\documentclass{article}

\usepackage{PRIMEarxiv}

\usepackage[utf8]{inputenc} 
\usepackage[T1]{fontenc}    
\usepackage{hyperref}       
\usepackage{url}            
\usepackage{booktabs}       
\usepackage{amsfonts}       
\usepackage{nicefrac}       
\usepackage{microtype}      
\usepackage{lipsum}
\usepackage{fancyhdr}       
\usepackage{graphicx}       
\usepackage{float}

\graphicspath{{media/}}     

\title{Urban planning in a context of rapid urban growth
\thanks{\textit{\underline{Citation}}: 
\textbf{Fadda, Margherita. 2024 Urban Planning in a Context of Rapid Urban Growth. Preliminary version.}} 
}

\author{
  Margherita Fadda \\
  Sahel and West Africa Club, OECD \\
  margherita.fadda@oecd.org}
  
\begin{document}
\maketitle

\begin{abstract}
As the African continent continues to urbanise, cities are becoming increasingly central to the transformations of societies and economies. Many studies highlight the limits of urban planning in these cities, emphasising the high share of population living in slums and the low levels of services that reach neighbourhoods. Less attention is given to the urban planning activities that try to prevent or improve these conditions.
This analysis of urban plans illustrates that plans are more widespread than commonly thought. They also, for the large part, consider spatial growth. The low number of cities that grew within the projected boundaries of these plans is a symptom of numerous bottlenecks that constrain planning systems in these countries. Examples of these include the disregard of the full built-up areas at the time of the plan's approval and the missing link between the plans and the financial means allocated for its delivery. This article identifies opportunities to overcome these barriers such as flexible and adaptable urban plans that consider the entire built-up area of the agglomeration.
\end{abstract}

\keywords{African cities \and Urban planning \and Urban growth \and Urbanisation \and Urban plan}

\section{Introduction}

The rapid demographic growth in the African continent has been changing its geography for decades. As the continent continues to urbanise, cities will become increasingly central to the transformations of African societies and economies. Cities will be home to an additional 700 million people by 2050, which will settle both within and outside existing built-up areas (OECD/AfDB/UCLGA, forthcoming). Urbanisation represents an opportunity to develop economic activities and improve the wellbeing of its residents. For cities to play their role in the continent’s economic growth, they need to adequately host new urbanites and service newly urbanised areas.

Urban planning is a crucial tool that can guide cities to support and accompany new residents and associated economic activities. Research has shown how regulations on the urban environment can have long-lasting consequences on the urban fabric (Carlucci and Salvati, 2023). As urban populations continue to grow at a rapid pace, it is essential to identify bottlenecks that prevent cities to be more effective at tackling key issues such as unhealthy living conditions. Addressing these before populations settle can potentially avoid expenses for future inefficiencies linked to urban forms, such as the lack of public space and green areas (Anderson, Patiño Quinchía and Prieto-Curiel, 2022).

This article contributes to the growing literature on urban planning tools in the African continent and the obstacles the practice faces to accompany urban growth and steer it in a sustainable path. Previous studies have evaluated whether existing plans are put into practice. This is often done at a national scale, delving into the detail of urban plans. Huang (2018), for example, analyses in depth the extent to which urban plans were present and applied in seven cities in Tanzania. Other approaches restrict the analysis to one urban plan (Mabaso et al., 2015) or to several plans within the same city (Chissola, 2015). These studies identify shortcomings and issues with plans and national legislations, informing directly local action. Despite their usefulness, they often restrict the analysis to capitals or larger cities (Kamana, Radoine and Nyasulu, 2024) and leave no room for cross-country comparisons.

Research that touches on urban planning at a continental or cross-regional scale tends to focus on the issues faced by cities, such as low service provision, congestion, crowding and slums (Lall, Henderson and Venables, 2017). Continental scale research also addresses the topic of normative frameworks for urban planning in African countries, without dealing directly with their effectiveness (UN Habitat, African Planners Association, 2014; Zimunya and Chirisa, 2021). It focuses on the shortcomings of the planning systems as a whole, presenting limited evidence on single instrument, such as the urban plans, themselves. 

On these bases, the article was guided by three main questions: 1) Are all cities equipped with a plan? 2) Do plans consider urban expansion? 3) Do plans use an accurate starting point to determine urban growth?
The present study aims to fill this gap. It provides the first analysis at large scale of urban plans in the African continent. The research included 227 cities across all African countries. 131 plans across 47 countries were found through convenient sampling and analysed. To be able to compare cross country, the analysis restricted its focus to easily observable evidence of the urban planning system, notably spatial urban plans. It starts from the hypothesis that a plan to steer growth needs to make room for urban growth. 

Other considerations, such as the street patterns or the respect of land uses are not considered. These elements are difficult to observe, they also rely on standards that vary widely city by city, thus compromising comparability. These approaches are based on assumptions that often ignore local contexts and risk importing rules and standards that do not fit local realities (Barnett and Parnell, 2016).

\section{Methodology}
\label{sec:headings}

To determine whether all cities selected are equipped with a plan, the first step was to identify a list of urban agglomerations. Definitions of a city vary greatly country by country, different criteria, such as population thresholds and levels of services make comparisons across countries impossible. This analysis uses a harmonised definition of urban areas determined and calculated for the continent by Africapolis. Africapolis identifies urban agglomerations using spatial satellite imagery, matched with census data. Urban agglomerations are those with 10~000 inhabitants or more, living in a continuously built-up form where buildings are at a maximum distance of 200 meters (OECD/SWAC, 2020). In 2015, Africapolis identified 7~600 urban agglomerations across the continent. The complete Africapolis list of cities was divided in three categories with equivalent size of population, resulting in three categories of cities: small, intermediary and large cities. The median population for one city in each category was respectively 1.6~million, 170~000 and 28~000.

\begin{table}[!htbp]
 \caption{Africapolis 2015 Data}

  \centering
  \begin{tabular}{lll}
    \toprule
    \cmidrule(r){1-2}
          & Population     & Number of Cities\\
    \midrule
    Large agglomerations & 123,890,180  & 54     \\
    Intermediary agglomerations     & 20,198,426 & 69      \\
    Small agglomerations     & 2,428,328      & 104  \\
  \end{tabular}
    
    \footnotesize
    
Note: The whole population of each country was divided in three categories with equivalent size of population. Source: (OECD/SWAC, 2020)
  \label{tab:table}
\end{table}

A cluster sampling methodology was used to identify a convenience sample of cities to look for plans. The research sought to collect 227 plans for agglomerations of all sizes and across the continent. Plans were looked for through desktop research. Common sources include official sources, published journal articles and PhD dissertations. The desktop research was conducted in English, French, Portuguese, Spanish and Arabic. Urban plans considered in the sample had to cover the urban municipalities they were designed for, this led to the exclusion of regional plans and neighbourhood layouts from the review.

For each of the 227 plans identified, the research sought to analyse a valid spatial representation of the city plans’ provisions. The research considered a spatial representation any geospatial map associated with the plan that referred to future urban development. Cities for which a spatial representation of a plan was found were analysed in depth. For some cities, the research identified evidence of the existence of a plan, but a digital copy could not be found. For the remaining cities, no plans could be found. In some cases, the non-existence of a plan was confirmed by policy papers or research that called for one to be developed. 

The spatial representation of the plans was analysed to determine whether they planned for growth. This could be indicated by explicit text included in the map legend, examples of names are "Terreno urbanizáveis" in Cubal, Angola, "Terres urbanisables" in Bujumbura, Burundi, "Zone d’urbanisation future" in Ziniare, Burkina Faso, "Proposed expansion" in Busia, Kenya. They could also determine them with their future land use in peripheral areas of the agglomeration ("High density residential zones" in Lilongwe, Malawi). In some cases, the legends were not clear enough to determine whether this was the case or not. In these instances, it was assumed that the plan considered future spatial growth if the plans had large, structured plots identified in the plan that were unbuilt when comparing with GHSL data.

The plans were categorised in two sets, the first one includes 62 plans for completion by 2020, which were prepared between 1974 and 2014, these are referred to as "past plans". The second one includes 69 plans for completion after 2020 (the latest one due for completion in 2050) and approved between 1997 and 2023, these are referred to as "current plans". Whilst the selection of 2020 as the cut-off year is arbitrary, several reasons motivate it: this date allows enough time post-plan completion (four years at the completion of this draft) which was deemed a necessary buffer to allow for its delivery.

To determine whether plans accurately planned for growth, they were compared with historical built-up area layers. The urban plans images were converted into .tif files, to be able to read them in a QGIS environment, and overlaid with Africapolis morphologies (year 2015), historical built-up data from the Global Human Settlement Layer (GHSL), which provides information globally from 1975 to 2020 (Pesaresi and Politis, 2023) and historical Google Earth imagery. The year of the plan’s approval determined the database used for the analysis.

This comparison determined:

a)	if the plan considered all the existing built-up area when approved, and

b)	the accuracy of the identified growth areas.

\subsection{Did the plan consider the existing built-up area when approved?}

A plan was classified as considering all the built-up area if all the GHSL, Africapolis or Google Imagery extents were covered by the plan at the time of the plan approval. In cases where the Africapolis and GHSL extents did not correspond, GHSL extents were chosen. This is because GHSL extents are available for many more dates (1975-2020) than Africapolis (2015- 20). The most recent available year available on GHSL was used and no control was made if the plan did not match the exact year (this was, however, not a common occurrence).

\begin{figure}[!htbp]
  \centering
  \includegraphics[width=0.6\textwidth]{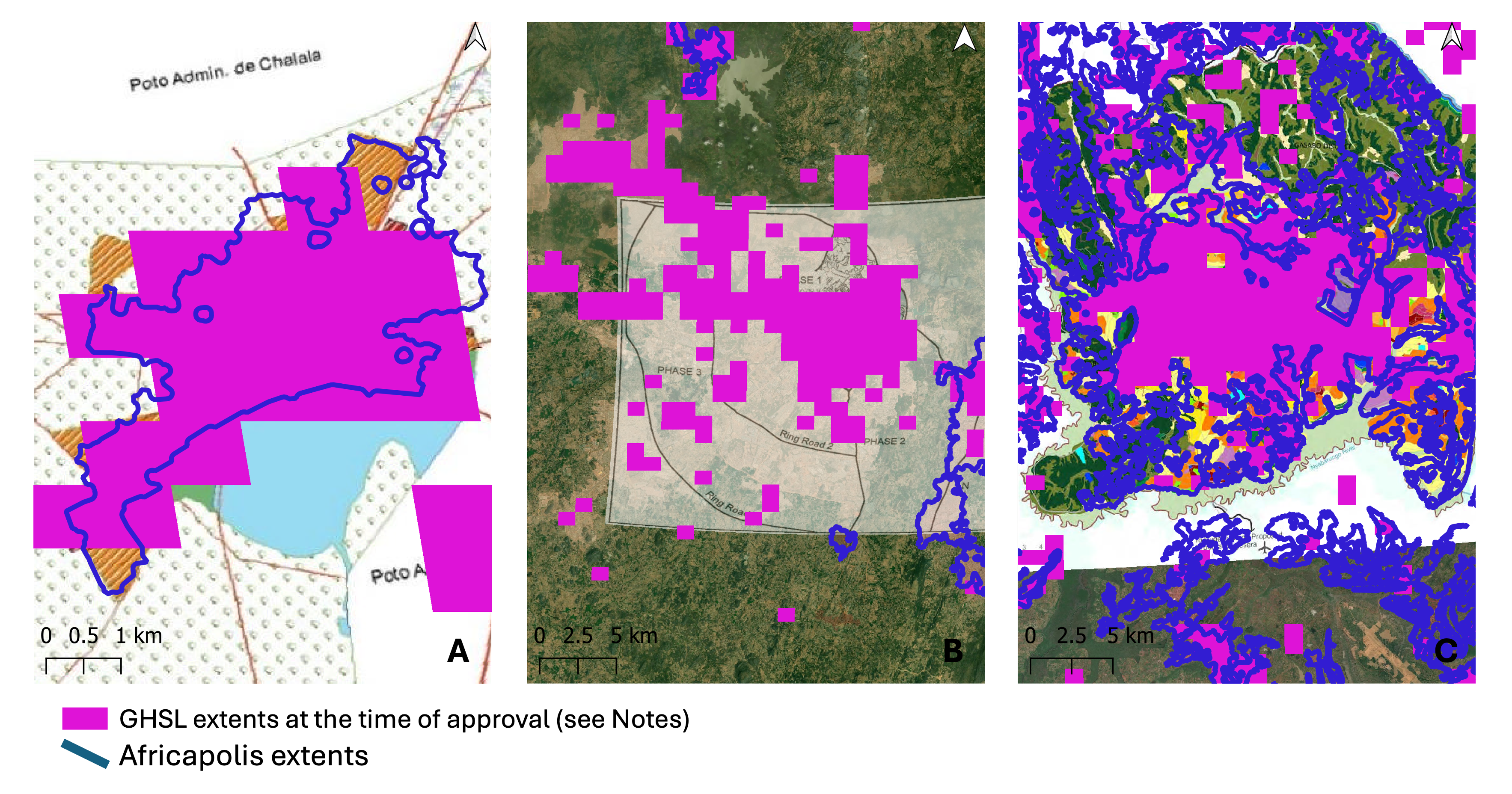}
  \caption{Location of the cities covered by the research}
  \label{fig:1}
  \medskip
  \footnotesize
    Note: A. Madlakaze, Mozambique, 2015 urban plan ; Example of a “yes”: no built-up extents outside the built-up area in 2015. Data source: Africapolis and GHSL (2015).  
    B. Abuja, Nigeria, 1979 urban plan ; Example of a “no”: built up extents outside of the plan (Northern areas). Data source: GHSL only available (1980).
    C. Kigali, Rwanda, 2018 urban plan ; Example where Africapolis and GHSL extents didn’t match. In consideration of GHSL data, the urban plan scores “yes”. No built-up extents outside the built-up area in 2018. Data source: Africapolis and GHSL (2018).
\end{figure}

\subsection{How accurately was growth planned for? }

Using the same built-up layers, the analysis determined whether the plan accurately planned for growth. The research revealed whether the growth areas were already built-up when the plan was approved and, for past plans, whether the growth happened only in the identified growth areas.

\subsubsection{Were growth areas identified in the plans already built-up when the plans got approved?}

By comparing the plan’s extents and the built-up extents at the time of the plan’s approval, it was possible to determine whether the growth areas identified in the urban plan were already occupied. In some cases, plans just display the land uses without making a difference between existing and future development areas. In these instances, if the built-up area in GHSL the year of the plan’s approval is smaller than the one outlined in the plan, the growth areas in the plan are assumed to be vacant. If the built-up area in the GHSL is built-up on the year of the plan’s approval, it was assumed that the plan's growth areas were built-up.

\subsubsection{Did the growth happen only in the identified growth areas?}

For past plans that planned for growth, the research verified whether the growth happened beyond the originally identified areas. If the plan considered future spatial growth, these areas were compared to the GHSL extents to verify whether the growth had extended beyond the outlined growth areas. The cases where it appeared that no growth had taken place were verified using Google Earth Pro imagery. If the plan did not consider future spatial growth, the research verified whether the growth happened only in the originally delimited area.

\section{Results and discussion}

The analysis provides rich insights into the status of urban planning across the continent. The limited amount of data that would normally allow to verify the conformance to the plan’s provisions – the effective reality on the ground compared to the visions proposed in the plans, such as land uses, number of households, etc. – limits the ability to evaluate in full the plans. Therefore, the results  estimate the ability of the plan to influence the city's growth by analysing three elements:

\begin{enumerate}
    \item Whether the plan existed or not
    \item The provision of areas for urban growth
    \item The accuracy of the projections of urban growth (for past plans only)
\end{enumerate}

\subsection{The existence of a plan}

The existence of these plans and their availability through desktop research is testament of a wide and rich field of urban planning across the continent. The research found 131 plans out of the 227 plans identified through the cluster sampling technique. An additional 29 plans from the original sample existed but were not found and therefore could not be analysed.  The remaining plans were categorised in two sets, depending on their year of completion, as explained in the methodology section. The cities for which a plan was found are shown in Figure 2. The complete list of agglomerations used in this research can be found in the Annex.

A plan existed for 70\% of the agglomerations searched for, however, the share of plans for which a spatial representation was available was 58\%. Plans remain most often an attribute of larger cities. Almost all larger cities have a plan, 93\%, the rate diminishes as the population drops: 74\% and 57\% of intermediary and smaller cities, respectively, possess plans (Figure 3).

Despite the numerous plans found - approximately 60\% of the original sample of cities - many cities still lack an urban plan. Approximately 40\% of the cities selected miss a plan. It is rare that larger cities and capitals miss a plan, this is the case for countries with very unstable political circumstances (Liberia, Somalia, and Eritrea). Smaller cities rarely have a plan. On the one hand, these cities might be missing the authority to prepare a plan at all. This is the case, for example, of localities that are not recognised as cities in their national legislation (this might be because of their population levels or missing characteristics) (Moriconi-Ebrard, Harre and Heinrigs, 2016). On the other hand, these cities – as well as intermediary ones – are endowed with less revenues than their larger counterparts. Budgets might in these cases be prioritising other spending items (AfDB/OECD/UNDP, 2016).
\begin{figure}[H]
  \centering
  \includegraphics[width=0.6\textwidth]{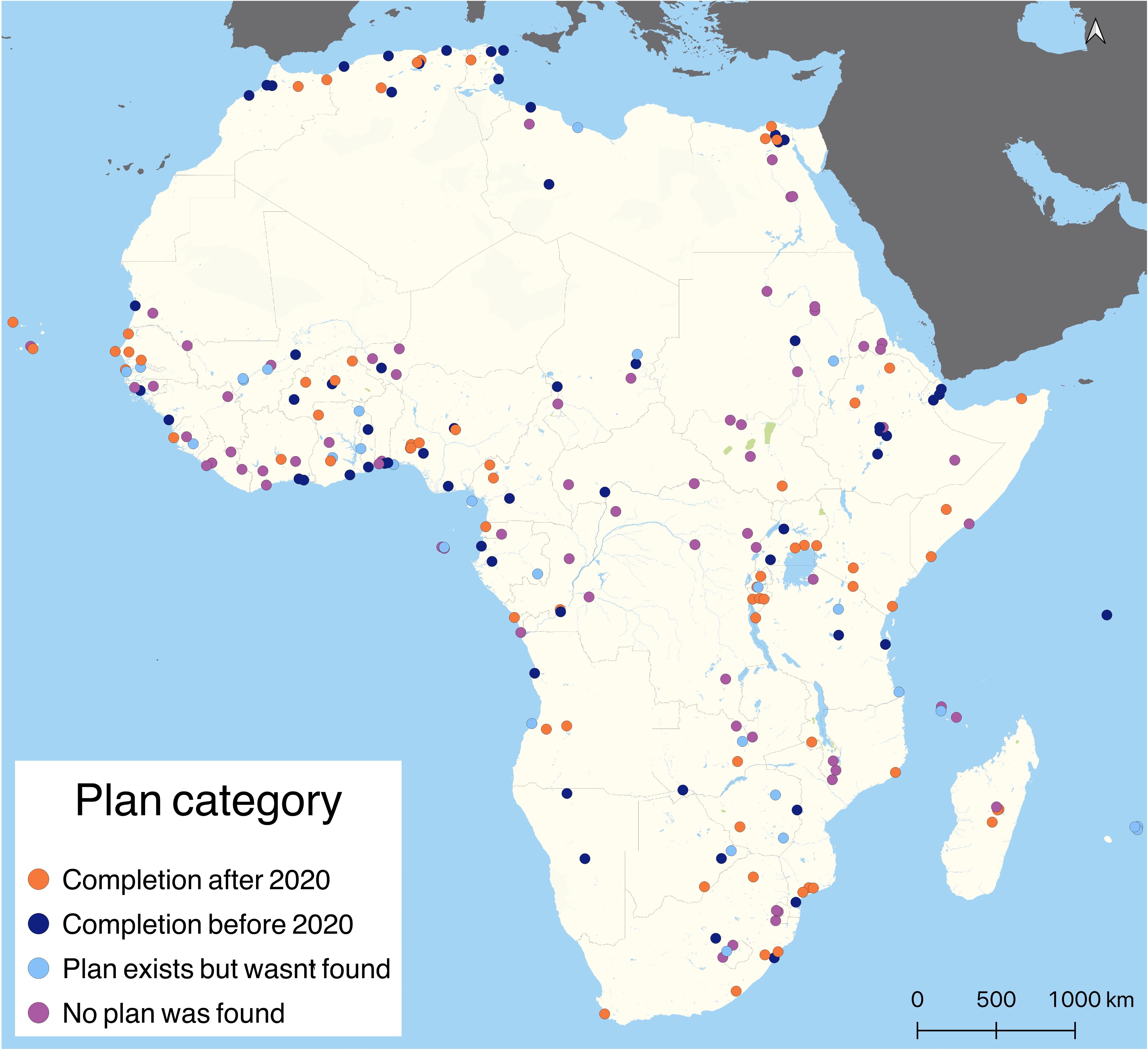}
  \caption{Location of the cities covered by the research}
  \label{fig:2}
  \medskip
  \footnotesize
    Note: Orange dots indicate cities with a plan for completion after 2020, blue dots indicate cities with plans for completion before 2020. Light blue dots indicate cities for which a plan exists but could not be found through desktop research. Purple dots indicate cities for which no evidence of a plan could be found. The research identified 131 cities for which a plan existed, 29 cities had a plan but it could not be found. For 67 cities no plan could be found.
    Source: Author and Africapolis (OECD/SWAC, 2020). The complete list of agglomeration and source reference for the plan can be found in the Annex. The referenced plans can be downloaded in a .tif version at www.mapwarper.net.
\end{figure}

\begin{figure}[!h]
  \centering
  \includegraphics[width=0.6\textwidth]{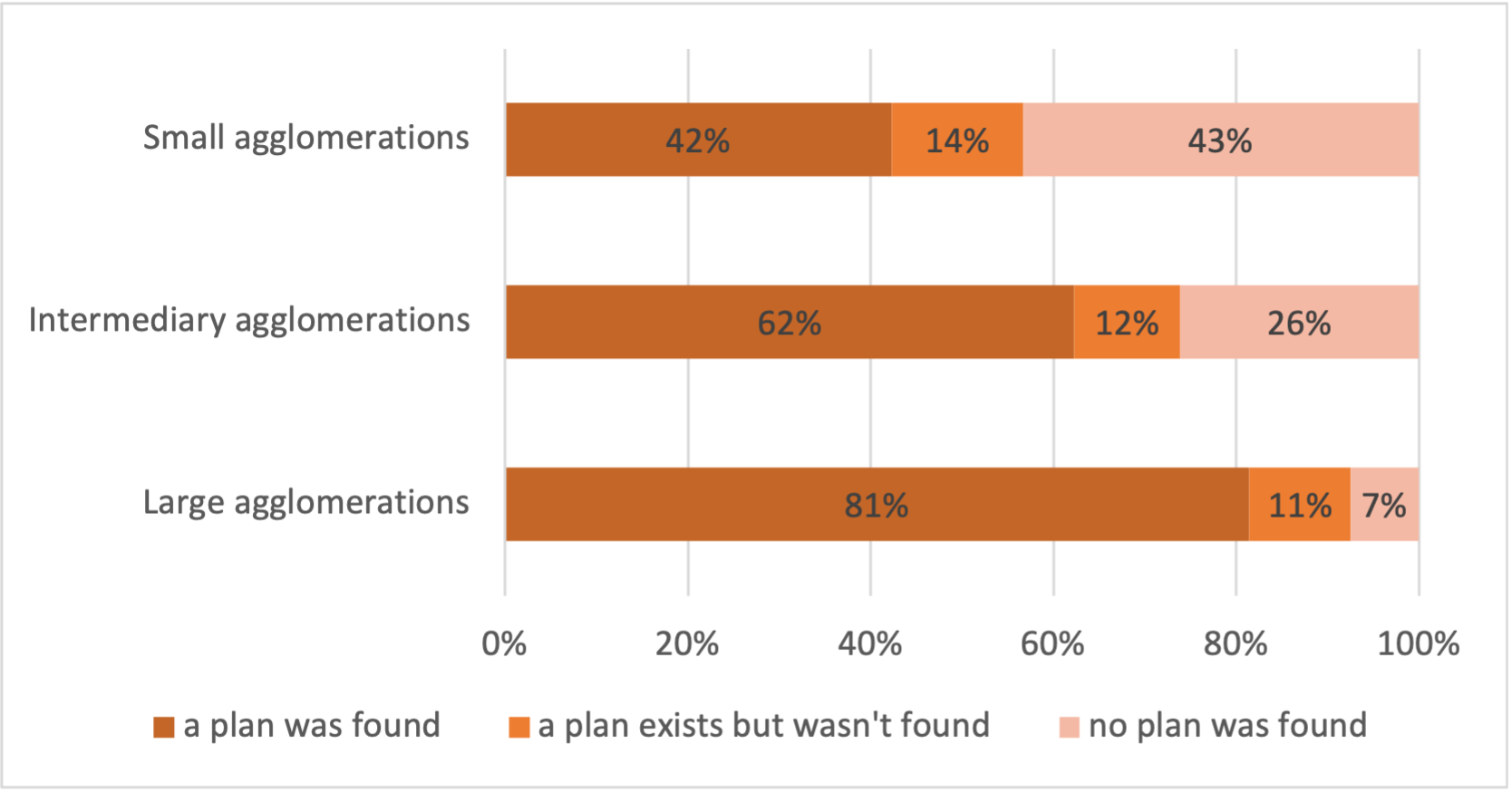}
  \caption{Plan status by city size}
  \label{fig:3}
  \medskip
  \footnotesize
    Note: Large agglomerations are more likely than intermediary and smaller ones to have an urban plan. Small agglomerations are the least likely to have a plan. Overall, 44 plans for large agglomerations, 43 plans for intermediary agglomerations and 44 plans for small agglomerations were analysed in this review.
\end{figure}

The share of plans found varies greatly across regions. The region for which the least number of plans was found was Central Africa: only 15 appeared to exist and 12 were found of a total sample of 31 plans. Over half of the plans selected for review weren’t found. The best performing region was Northern Africa, for which only 22\% of the plans selected for review weren’t found (Figure 4). Explaining these regional differences is outside of the scope of the research but possible explanations that could be pursued further include the level of urbanisation of these regions, their wealth and urban policies. Central Africa is also the region with the least approved national spatial strategies and urban strategies, showing a lag in the thinking about urbanisation.

\begin{figure}[!htbp]
  \centering
  \includegraphics[width=0.6\textwidth]{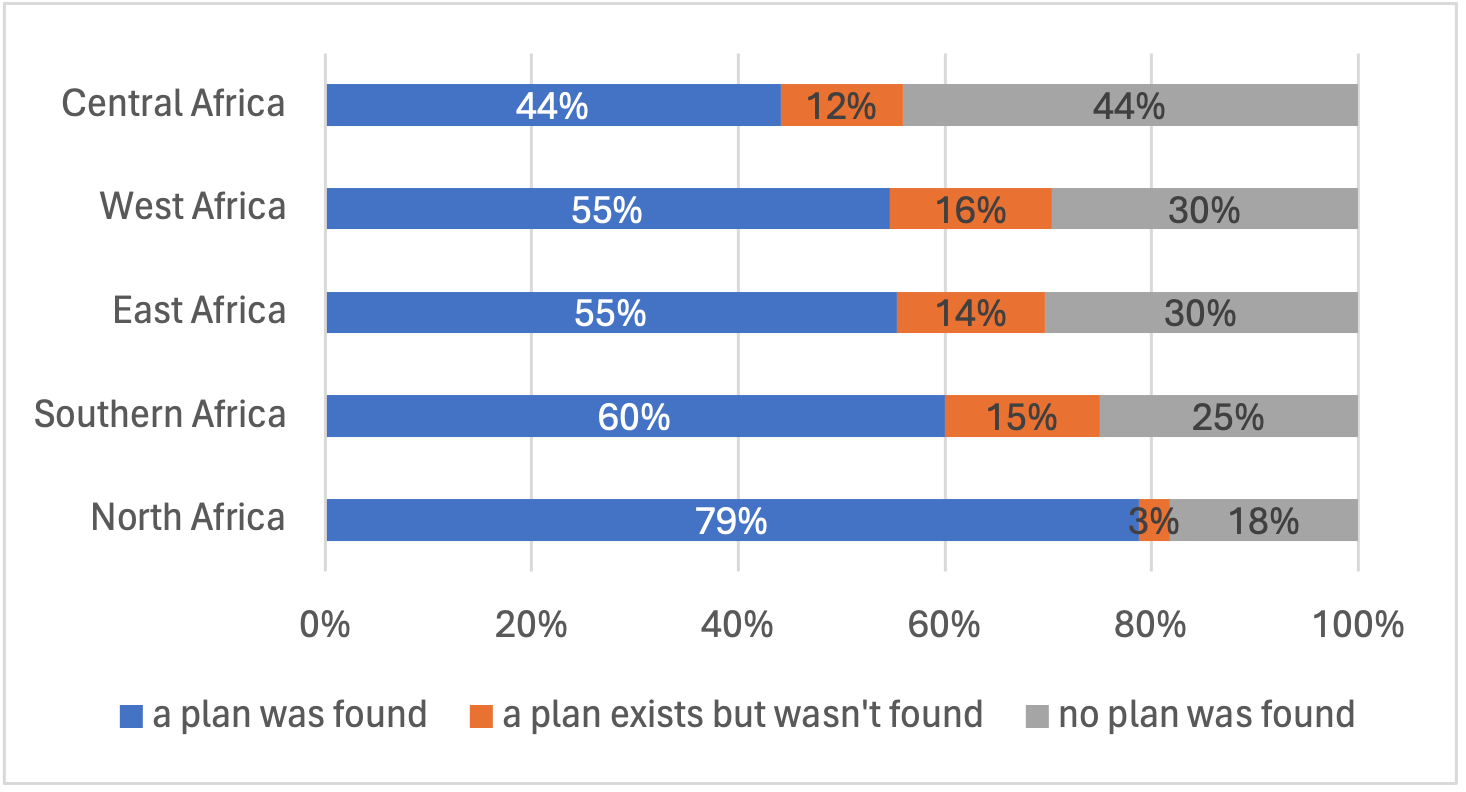}
  \caption{Plan status by region}
  \label{fig:4}
  \medskip
  \footnotesize
    Note: Percentages obtained dividing the total number of plans for the region and the number of plans in the database. Northern Africa is the region for which evidence of a plan and their spatial representation were found (82\%).
\end{figure}

\subsection{Plans’ provisions for growth}

Across the 131 plans for which a spatial representation was available, 121 included areas for growth (92\%). The presence of growth areas in such plans indicates an awareness of municipalities and urban planning officers of the phenomenon of urban growth. Past plans for large agglomerations were more likely than current ones to plan for growth, 59\% of them did so in the past, against just 30\% of current plans. Inversely, current plans for small agglomerations are more likely than past ones to plan for growth, with 66\% and 30\% of them respectively doing so.

Despite the prevalence of growth plans, urban plans appear to have often underestimated spatial growth over time. Only 18\% of all past plans, 11 out of 62, grew within the limits of projected spatial expansion. 

\begin{figure}[!htbp]
  \centering
  \includegraphics[width=0.6\textwidth]{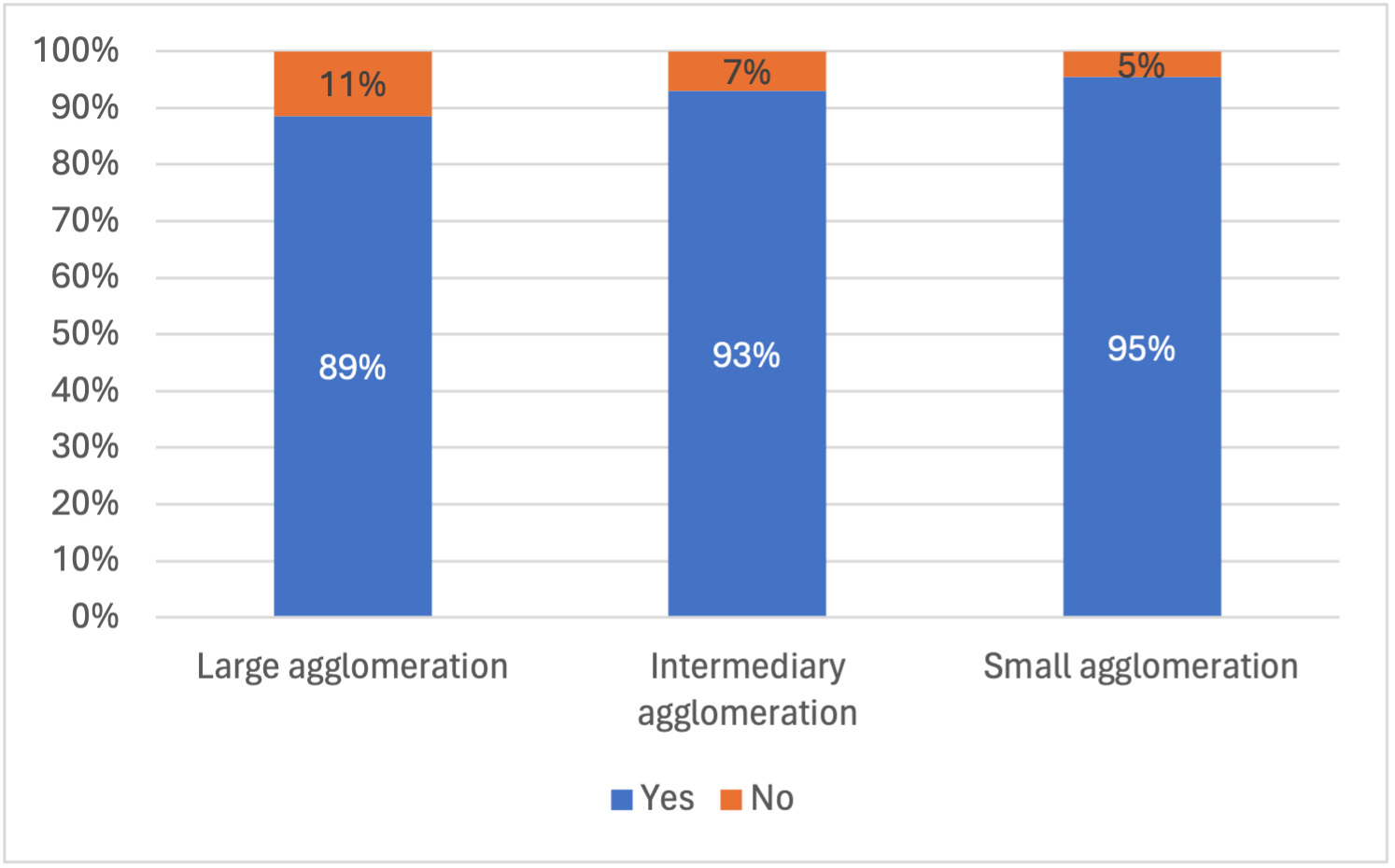}
  \caption{Consideration of urban growth by city size}
  \label{fig:5}
  \medskip
  \footnotesize
    Note: Most cities analysed planned for growth. Large agglomerations are slightly less likely than intermediary and smaller ones to plan for growth.
\end{figure}

\subsection{Results for accurate identification of the city footprint}

Many reasons could explain why plans inadequately projected spatial growth, with obstacles both at the plan preparation and implementation stage. The lack of accurate assumptions of population and spatial growth could explain this shortcoming. Because of the limited availability of the full planning documents that accompany these plans, it is very difficult to understand the part of responsibility of the plans’ projections. For those where this documentation was available, demographic projections have matched or overestimated the observed growth, but spatial requirements linked to it have not been mentioned or were underestimated. 

Two potential explanations for the misestimates of expansion include (1) misrepresentation of the starting point, i.e., true built-up extent and (2) the inaccurate projection of the direction in urban growth. Both of these explanations concern the plan preparation stage, as plans did not use adequate data for urban footprint, spatial attractors for growth or were constrained by administrative boundaries. Implementation shortcomings might also be responsible. The next section explores these obstacles in more depth.

\subsubsection{Misrepresenting the starting point}

Misrepresenting the starting point at the time of plan’s approval is a key reason for inaccurately projecting expansion. Because spatial plans are available for the complete sample of plans analysed, their coverage can easily be compared to the extents of the cities at the time the plans were approved. As outlined in the methodology section, plans were compared to various sources of satellite data to determine the accuracy with which plans considered the entire agglomeration extents.

This comparison reveals that 34\% of all plans did not consider the whole built-up extents at the time of approval. "Past plans", plans for completion before 2020, are more likely to disregard built-up extents, 48\% of them do so, compared to 20\% of current plans (Figure 6, left). Figure 7A illustrates an example of a plan that grew beyond the initially identified areas of growth. A common explanation for this shortcoming lies in the misalignment between the administrative city and the effective built-up area. Determining which neighbourhoods are generally excluded from administrative boundaries and therefore urban plans is beyond the scope of this paper, however, recent research shows that in informal settlements are more likely to be excluded from these boundaries (Boanada-Fuchs, Kuffer and Samper, 2024).

Smaller cities are the most likely ones to consider their full built-up extents, 84\% of them do, compared to 72\% of intermediary cities and 43\% of larger cities (Figure 4, right). Current plans for all city sizes are more likely to consider the full built-up extents. Smaller cities fare better in this category too, with 87\% of their plans considering the whole built up extent, against 73\% of intermediary cities and 75\% of larger cities. Figure 7B illustrates an example of a more recent plan of a small city that considers the full built-up extents. Current plans for larger cities are much more likely than before to consider full built-up extents, only 25\% of them did in past plans, compared to 75\% in current plans (Figure 6, right). 

\begin{figure}[!htbp]
  \centering
  \includegraphics[width=0.6\textwidth]{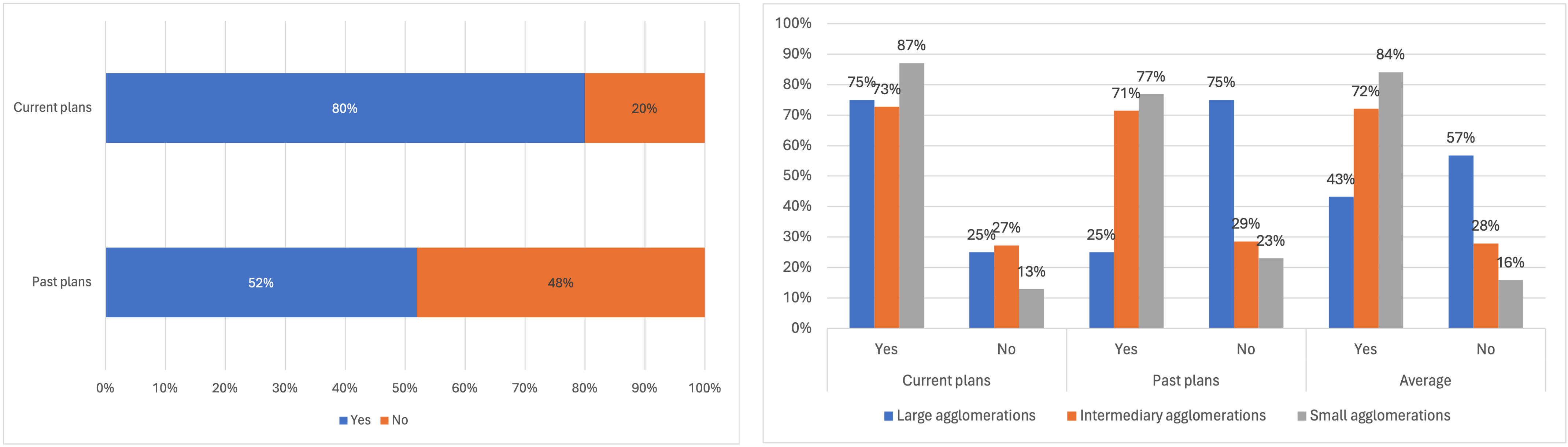}
  \caption{Did the plan consider all the built-up area at the time of approval?}
  \label{fig:6}
  \medskip
  \footnotesize
    Note : These graphs show the share of plans that consider the entire built-up extents at the time of their approval. The plans are grouped in two categories, those for completion before 2020 (past plans), and those for completion after 2020 (current plans). The graph on the left shows the general trend that current plans are more likely than past plans to consider the entire built-up area. The graph on the right shows how plans for differently sized cities consider actual city extents. Plans for small agglomerations are overall more likely to consider the entire built-up area than those for intermediary and large agglomerations.
\end{figure}

Current plans have likely benefitted from the use of new technologies, particularly in the field of geographic information systems (remote sensing, satellite imagery, geo-localised surveys etc., collaborative map making platforms). These improvements have made it easier to identify challenges cities face and the resources at their disposal to establish the objectives of the plan. The changing consideration of informal areas, once excluded from official planning, might have also influenced the ability of considering these areas in the plan (Friedrich Ebert Stiftung, 2022).

Different size cities benefit from these advances to varying degrees. Intermediary and small cities seem to benefit from improvements in the conditions of plan-making. They have a wider window of opportunity to structure new sectors of the city before encountering spatial geography limitations or administrative boundaries that could compromise their plans for growth (Figure 7C). This result is tempered by the limited reforms in the field of local governance that will impact the growth if and when these agglomerations reach the existing administrative boundaries (Zimbabwe, 2019).

Larger cities, despite have more up-to-date plans, are less likely to exploit the improvements mentioned above. Similarly to the previous generation of plans, the full extents of the built-up area are often ignored. Administrative boundaries that have seldom changed since the 1990s limitations (Figure 7C) can at times explain this. Current plans with completion date up to 2050 bet on densification as a measure to host new inhabitants (ex. Banjul, Freetown), turning a blind eye to where the spatial growth is taking place.

\begin{figure}[!htbp]
  \centering
  \includegraphics[width=0.6\textwidth]{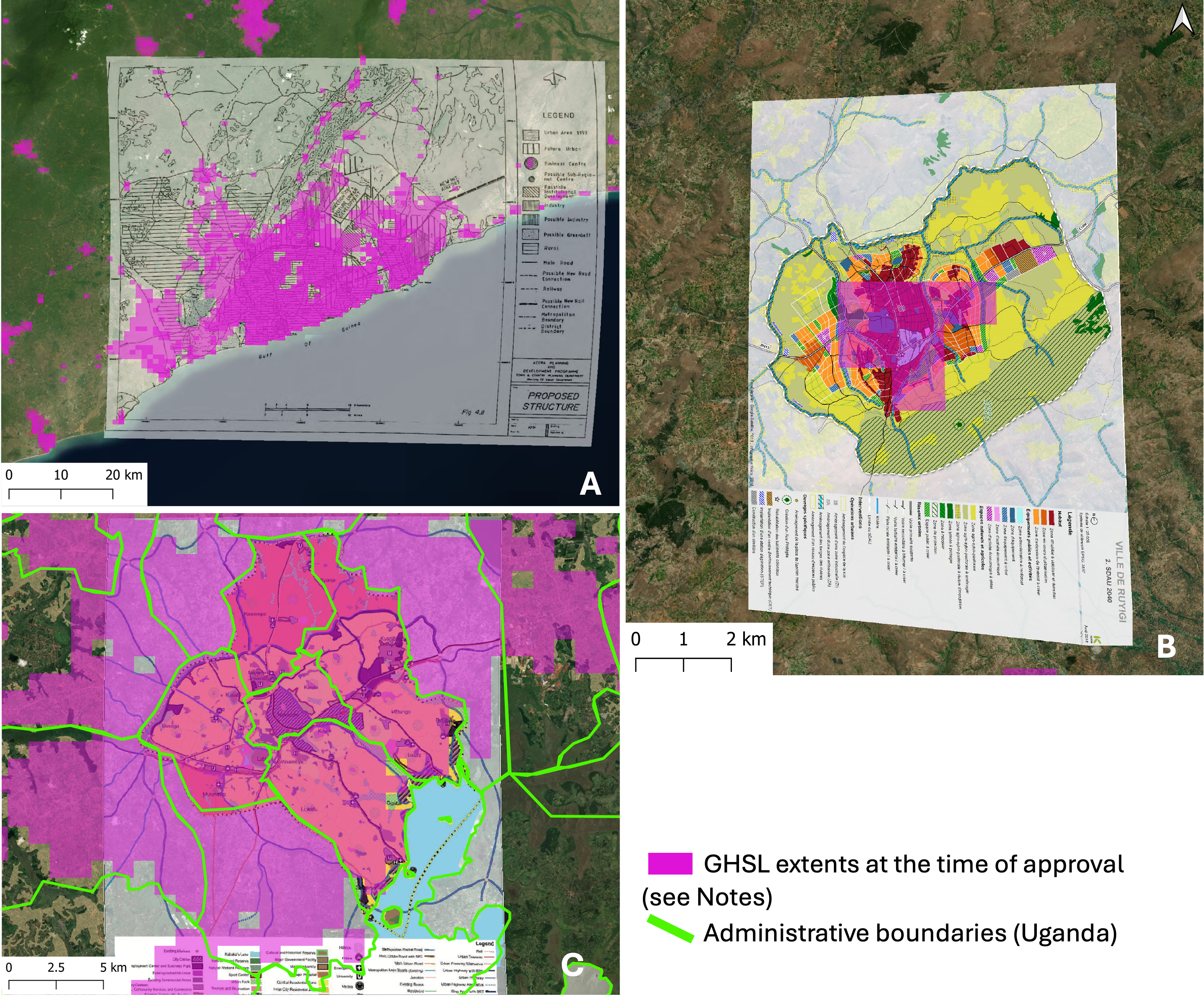}
  \caption{Urban Plans}
  \label{fig:7}
  \medskip
  \footnotesize
    Note : 7A. Strategic Plan for the Greater Accra Metropolitan Area 1992-2010 (Ghana) ; An example of a plan that disregards the full built-up extents at the time of approval (Ministry of Local Government, 1992); (Pesaresi and Politis, 2023) ; 7B. Schéma directeur d’aménagement et d’urbanisme de Ruyigi 2018 -2040 (Ville de Ruyigi/Keios, 2018[18]) (Pesaresi and Politis, 2023); 7C. Kampala Physical Development Plan 2012-40 (Uganda); (Pesaresi and Politis, 2023) and Kampala Capital City Authority (2012).
\end{figure}

\subsubsection{The direction of growth}

In some cases, the growth did not happen within the area of the plan at all, but in other directions. The lack of implementation of parts of the plan, such as a road or other infrastructure, can explain this. These circumstances make the area originally identified for growth impracticable or less desirable than other areas already available. The example of Banoua, Côte d’Ivoire, illustrates this. A missing road network and the lack of a detailed implementation plan resulted in the inability to deliver the plan and spatial growth outside of the plan’s perimeter. Successive analysis identified the delays and failure to deliver the road link to this new area as the reason for its failure of developing as expected (MCLAU, JICA, 2015, p. 421).

This might also explain the growth patterns in Casablanca, Morocco. In its 1985 plan, the city planned to develop linearly, along the connection between the capital and Mohammedia. The city developed concentrically instead, disregarding the patterns of development indicated in the plan. The missing links between the urban hubs were identified as the reason behind the failure of the linear development in the plan’s evaluation report:
«The “Primary Road Network” plan of the 1985 SDAU (Schéma Directeur d’Aménagement Urbain; Urban Development Master Plan) compared to the current situation highlights the following:  the failure to build the roads between Casablanca and Mohammedia resulted in the failure to implement the linear development option of the agglomeration (...)» (Agence Urbaine de Casablanca, 2000, p. 33).

\subsection{Conditions}
The conditions under which these plans were prepared can also explain their shortcomings predicting and directing growth.

\subsubsection{The time of approval}
When looking at the plan-making procedures, long duration of the drafting or approval periods of such plans can explain inadequate and outdated plans by the time of approval. An infamous example of this is the Schéma directeur d’aménagement du grand Tunis, finalised in 1995 and approved in 2004 (Chabbi and Abid, 2008). Over the same period, the 315 000 new residents of Tunis (OECD/SWAC, 2020) settled in an agglomeration without a clear plan in place. The lapse of time between the initial analysis carried out to make the plan (diagnosis), its completion and approval also means that the built-up extent considered in the plan will unlikely respect the truth encountered on the ground at the implementation stage. The long period required for the completion of these plans was and still is often linked to their complexity. The requirement for extensive data on all aspects of the agglomeration is often a counterproductive exercise that determines the inability of the plan to deliver a practical structure to the city at the time when it’s needed. Since the late 1990s professionals in the urban development field argued for simpler plans, sometimes called “Reference plans” (Farvacque-Vitkovic and Godin, 1997).

\subsubsection{The involvement of local actors}
Often led by central governments in tandem with development agencies  and foreign planners, the plan making left little room of manoeuvre to local actors on the ground, be it local administration or communities, to contribute and inform plans. The plans have been criticised for being blind to the existing logics of communities settled, notably land politics and ownerships (Kanyiehamba, 1974; ISTED, 1998; UN-Habitat, 2009). These shortcomings have sometimes played a crucial role in undermining the delivery of plans whose objectives didn’t align or contradicted those of local stakeholders excluded from the planning procedures. The exclusion of these actors at the plan making stage also determined a missed opportunity for knowledge transfer to planners and officers in charge of implementing these same plans (Farvacque-Vitkovic and Godin, 1997). The current data available through this review does not allow to verify whether more recent plans might be benefitting from more local know-how. Nevertheless, the growing number of urban planning schools in the continent and increasingly vocal critics of imported models are hopeful signs that this is changing. The improvements in the consideration of the full built-up extents might be a result of this. 

\subsubsection{The budget problem}
Limited financing sources allocated for the delivery of urban plans also compromised their success. On the one hand, low finances at the municipal level translated in weak planning regulations enforcement. This led to the lack of enforcement of the plans rules (parcels, no-aedificandi, etc.). On the other hand, the missing link between plans objectives in terms of infrastructure development and funding available compromised the ability to deliver them. The older generation of plans drafted in the 1980s and 1990s were particularly impacted by structural adjustments, which reduced the means available to public administration. The progressive disengagement of state actors from the practice developed a domain almost entirely reliant on international donors and expertise (Kanyiehamba, 1974; Harrison and Croese, 2022).

Existing urban plans often do not have a clear path for implementation and plans for financial and human resources management. Some plans that address the financing question limit themselves to a paragraph mentioning the requirement for funding to be allocated for the delivery of the infrastructure called for in the plan. Development partners often finance these plans, without the consistent involvement of local governments. Without this involvement, plans do not become a local priority and become irrelevant once the development partner project ends. Critics also point out that plans are often too ambitious and do not adequately consider the means at disposal. Simpler plans that consider the sequencing of investments and the prioritisations of infrastructure delivery are often raised as potential solutions to lower the costs of implementation (Lamson-Hall et al., 2019).

\section{Conclusion}
The review of urban plans illustrated in this article showed that plans are more widespread than commonly thought. They also, for the large part, consider growth. The low number of cities that grew within the boundaries of these plans is a symptom of numerous bottlenecks that constrain planning systems in these countries. Examples of these include the disregard of the full built-up areas at the time of the plan’s approval and the missing link between the plans and the financial means allocated for their delivery.  

As illustrated in the discussion, plans have improved their ability to consider their entire built-up extents. It is still too early to judge whether these improvements are enough to lead to more performing plans. The insights on the bottlenecks that determined the failure of past plans can help forecast their likely results and challenges they need to overcome - those related to the plan itself and to the planning systems. 
Technical tools might further contribute to improve the plans’ ability to adequately quantify spatial growth (Korah, Koch and Wimberly, 2024; OECD/AfDB/UCLGA, forthcoming). For these estimates to be successfully implemented and translated in results on the ground – planning systems need to be adapted. Two improvements seem to be a priority considering this review. 

Firstly, planning systems need to be able to be flexible and plan at the right scale, tackling administrative boundaries that prevent collaboration between jurisdictions. Without such improvement, plans will keep ignoring the areas where spatial growth it’s taking place. Secondly, plans need to calibrate their objectives to available resources. This is a complex and politically costly operation, however, a needed one to ensure structured growth and the long-term preservation of urban planning objectives (such as an efficient mobility system and adequate provision for parks and public offices).

Future research could build on these findings to explore in more detail the obstacles cities encounter to prepare an urban plan. Country-level analysis could provide more information on the limits and required reforms to build more effective planning systems. Urban planning is a key tool to manage the rapid urban growth experienced by African countries, leveraging all available resources can bring about plans that effectively steer urban development.

\section*{Acknowledgments}
I am grateful for the support of the Sahel and West Africa Club (SWAC/OECD), and in patricular to Philipp Heinrigs, Brilé Anderson, Inhoi Heo and Patrick Lamson-Hall.

\section*{Bibliography}
\bibliographystyle{unsrt}  

Abubakari, A. and D. Romanus (2011), “Urbanisation and the challenges of development controls in Ghana: A case study of Wa Township”, Journal of Sustainable Development in Africa, Vol. 13/07, p. 26, \url{https://jsd-africa.com/Jsda/Vol13No7-Winter2011A/PDF/Urbanisation%20and%20the%20Challenges%20of%20Development.Abubakari%20Ahmed.pdf}

AfDB/OECD/UNDP (2016), African Economic Outlook 2016: Sustainable Cities and Structural Transformation, OECD Publishing, Paris, https://doi.org/10.1787/aeo-2016-en.
Agence Urbaine de Casablanca (2000), Évaluation Prospective du SDAU de 1985 de la Région du Grand Casablanca, \url{https://auc.ma/pdf-reader/?dataURL=https://auc.ma/wp-content/uploads/2020/11/Eavluation-prospective-du-SDAU-1985.pdf#dearflip-df_4057/3/.}

Anderson, B., J. Patiño Quinchía and R. Prieto-Curiel (2022), Boosting African cities’ resilience to climate change, OECD Publishing \url{https://www.oecd-ilibrary.org/docserver/3303cfb3-en.pdf?expires=1712855089&id=id&accname=ocid84004878&checksum=21AAB7786CB3FA408595EBE1C53821F3}.

Barnett, C. and S. Parnell (2016), “Ideas, implementation and indicators: epistemologies of the post-2015 urban agenda”, Environment and Urbanization, Vol. 28/1, pp. 87-98, \url{https://doi.org/10.1177/0956247815621473}.

Boanada-Fuchs, A., M. Kuffer and J. Samper (2024), “A Global Estimate of the Size and Location of Informal Settlements”, Urban Science, Vol. 8/1, p. 18, \url{ttps://doi.org/10.3390/urbansci8010018}.

Carlucci, M. and L. Salvati (2023), “Assessing path-dependent urban growth with geographically weighted regressions”, Environmental Impact Assessment Review, Vol. 98, p. 106920, \url{https://doi.org/10.1016/j.eiar.2022.106920}.

Chabbi, M. and H. Abid (2008), La mobilité urbaine dans le Grand Tunis. Évolutions et perspectives, \url{https://planbleu.org/wp-content/uploads/2008/05/rapport_mobilite_urbainetunis.pdf}.

Chant, S. (2013), “Cities through a “gender lens”: a golden “urban age” for women in the global South?”, Environment and Urbanization, Vol. 25/1, pp. 9-29, \url{https://doi.org/10.1177/0956247813477809}.

Chissola, A. (2015), A Influência do processo de planeamento e gestão. O caso de estudo da cidade de Luanda, \url{https://www.google.com/url?sa=t&rct=j&q=&esrc=s&source=web&cd=&cad=rja&uact=8&ved=2ahUKEwj1oYzNg8KDAxVJT6QEHaV5Ay8QFnoECBMQAQ&url=https%3A%2F%2Ffenix.tecnico.ulisboa.pt%2FdownloadFile%2F563345090414837%2FAdilson%2520A.%2520A.%2520Chissola%2C%252079470.pdf}.

Coquery-Vidrovitch, C. (1995), “Histoire de l’urbanisation africaine”, in Panoramas urbains, ENS Éditions, \url{https://doi.org/10.4000/books.enseditions.25963}.

Dodman, D. et al. (2023), “Cities, Settlements and Key Infrastructure”, in Climate Change 2022 – Impacts, Adaptation and Vulnerability, Cambridge University Press, \url{https://doi.org/10.1017/9781009325844.008}.

Farvacque-Vitkovic, C. and L. Godin (1997), L’avenir des villes africaines, Enjeux et priorités du développement urbain, \url{https://invenio.unidep.org/invenio/record/4621/files/godin.pdf}.

Friedrich Ebert Stiftung (2022), Just City in Africa. The Transformative Value of Urbanisation, \url{https://library.fes.de/pdf-files/bueros/kenia/19697.pdf}.

Harrison, P. and S. Croese (2022), “The persistence and rise of master planning in urban Africa: transnational circuits and local ambitions”, Planning Perspectives, Vol. 38/1, pp. 25-47, \url{https://doi.org/10.1080/02665433.2022.2053880}.

Huang, C. et al. (2018), Translating Plans to Development: Impact and Effectiveness of Urban Planning in Tanzania Secondary, World Bank.

Ikiriko, T. and M. Udom (2023), “Staffing and resource capacity for effective implementation of master plans: a case study of Greater Port Harcourt City Development Authority in Port Harcourt, Nigeria”, International REsearch Journal of Modernization in Engineering Technology and Science, Vol. 05/04, p. 10, \url{https://www.irjmets.com/uploadedfiles/paper/issue_4_april_2023/35411/final/fin_irjmets1680844811.pdf}.

ISTED (1998), Dynamiques de l’urbanisation de l’Afrique au sud du Sahara.
Kamana, A., H. Radoine and C. Nyasulu (2024), “Urban challenges and strategies in African cities – A systematic literature review”, City and Environment Interactions, Vol. 21, p. 100132, \url{https://doi.org/10.1016/j.cacint.2023.100132}.

Kanyiehamba, G. (1974), “Urban Planning Law in East Africa”, Progress in Planning, Vol. 2, pp. 1-83, \url{https://doi.org/10.1016/0305-9006(74)90005-1}.

Korah, A., J. Koch and M. Wimberly (2024), “Understanding urban growth modeling in Africa: Dynamics, drivers, and challenges”, Cities, Vol. 146, p. 104734, \url{https://doi.org/10.1016/j.cities.2023.104734}.

Lall, S., J. Henderson and A. Venables (2017), African Cities: Opening doors to the world, World Bank.

Lamson-Hall, P. et al. (2019), “A new plan for African cities: The Ethiopia Urban Expansion Initiative”, Urban Studies, Vol. 56/6, pp. \url{https://doi.org/10.1177/0042098018757601}.

Mabaso, A. et al. (2015), “Urban physical development and master planning”, Journal for Studies in Humanities and Social Sciences, ISSN 2026-7215, pp. 72-88, \url{https://www.researchgate.net/publication/282294678_Urban_physical_development_and_master_planning_in_Zimbabwe_An_assessment_of_conformance_in_the_City_of_Mutare}.

Macharia, P. et al. (2023), “Exploring the urban gradient in population health: insights from satellite-derived urbanicity classes across multiple countries and years in sub-Saharan Africa”, BMJ Global Health, Vol. 8/10, p. e013471, \url{https://doi.org/10.1136/bmjgh-2023-013471}.

MCLAU, JICA (2015), Le projet de développement du schéma directeur d’urbanisme du Grand Abidjan (SDUGA). Rapport Final.

Ministry of Local Government (1992), Strategic Plan for the Greater Accra Metropolitan Area, \url{http://mci.ei.columbia.edu/files/2013/03/AMA-Strategic-Plan-vol-1.pdf}.

Molina, J. (ed.) (2023), “Structural transformation and the gender pay gap in Sub-Saharan Africa”, PLOS ONE, Vol. 18/4, p. e0278188, \url{https://doi.org/10.1371/journal.pone.0278188}.

Moriconi-Ebrard, F., D. Harre and P. Heinrigs (2016), Urbanisation Dynamics in West Africa 1950–2010: Africapolis I, 2015 Update, West African Studies, OECD Publishing, Paris, \url{https://dx.doi.org/10.1787/9789264252233-en}.

OECD/AfDB/UCLGA (forthcoming), Africa’s Urbanisation Dynamics 2024. Planning for Urban Expansion.

OECD/SWAC (2023), Africapolis (database), \url{http://www.africapolis.org/} (accessed on 19 October 2023).

OECD/SWAC (2020), Africapolis (database), \url{https://africapolis.org/} (accessed on 2 September 2022).

OECD/SWAC (2020), Africa’s Urbanisation Dynamics 2020: Africapolis, Mapping a New Urban Geography, West African Studies, OECD Publishing, Paris, \url{https://doi.org/10.1787/b6bccb81-en}.

OECD/SWAC (2013), Settlement, Market and Food Security, West African Studies, OECD Publishing, Paris, \url{https://doi.org/10.1787/9789264187443-en}.

OECD/UN ECA/AfDB (2022), Africa’s Urbanisation Dynamics 2022: The Economic Power of Africa’s Cities, OECD Publishing, \url{https://doi.org/10.1787/3834ed5b-en}.

Onokerhoraye, A. (1975), “Urbanism as an organ of traditional African civilization : The example of Benin, Nigeria / L’URBANISME, INSTRUMENT DE LA CIVILISATION AFRICAINE TRADITIONNELLE : L’EXEMPLE DE BENIN, NIGERIA”, Civilisations, Vol. 25/3/4, pp. 294-306, \url{https://www.jstor.org/stable/41229293}.

Paulais, T. (2012), Financer les villes d’Afrique. L’enjeu de l’investissement local, The International Bank for Reconstruction and Development/ The World Bank, Washington, DC, and Agence Française de Développement, Paris.

Pesaresi, M. and P. Politis (2023), GHS-BUILT-S R2023A - GHS built-up surface grid, derived from Sentinel2 composite and Landsat, multitemporal (1975-2030), \url{https://doi.org/10.2905/9F06F36F-4B11-47EC-ABB0-4F8B7B1D72EA}.

Ryser, J. and T. Franchini (eds.) (2015), International Manual of Planning Practice, International Society of City and Regional Planners, ISOCARP.

Seager, J. and K. Toepfer (2005), “Gender, Environment and the Millennium Development Goals: the UNEP Perspective”, Perceptions, This paper was specially commissioned by the United Nations Environmnet Programme (UNEP) for its Global Environmnet Outlook GEO Year Book 2004/5, pp. 115-140, \url{http://sam.gov.tr/pdf/perceptions/Volume-X/summer-2005/Seager-Toepfer.pdf}.

UN Habitat, African Planners Association (2014), The State of Planning in Africa, An Overview, UN-Habitat, \url{https://unhabitat.org/the-state-of-planning-in-africa-an-overview}.

UNDESA (2022), World Population Prospects 2022, \url{https://population.un.org/wpp/Download/Standard/MostUsed/}.

UN-Habitat (2009), Global Report on Human Settlements 2009. Planning Sustainable Cities: Policy Directions.

UNICEF (2018), Shaping urbanization for children. A handbook on child-responsive urban planning, \url{https://www.unicef.org/media/47616/file/UNICEF_Shaping_urbanization_for_children_handbook_2018.pdf} (accessed on 5 October 2023).

Ville de Ruyigi/Keios (2018), Schéma directeur d’aménagement et d’urbanisme de Ruyigi 2018-2040, \url{https://www.keios.it/wp-content/uploads/2020/01/1701-20-Ruyigi-SDAU-2040-1.png}.

Watson, V. (2015), “The allure of ‘smart city’ rhetoric”, Dialogues in Human Geography, Vol. 5/1, pp. 36-39, \url{https://doi.org/10.1177/2043820614565868}.

Zimbabwe, T. (2019), Mutare city runs out of land, turns to corner stands, air pockets, \url{https://allafrica.com/stories/201909260632.html} (accessed on 29 July 2023).

Zimunya, W. and I. Chirisa (2021), “Urban Policy and the Future of Urban and Regional Planning in Africa”, in The Palgrave Encyclopedia of Urban and Regional Futures, Springer International Publishing, Cham, \url{https://doi.org/10.1007/978-3-030-51812-7_109-1}.

\end{document}